\documentclass[showpacs,preprintnumbers,nourl,jcp,superscriptaddress]{revtex4-1}
\usepackage{amsmath,amssymb}
\usepackage{graphicx,psfrag}
\usepackage{dcolumn}
\usepackage{bm,tabularx,ragged2e,booktabs,caption}
\usepackage{ulem}
\usepackage{amsmath}
\usepackage{setspace}
\usepackage{leftidx}
\usepackage[english]{babel}
\usepackage{color}
\usepackage{subcaption}
\usepackage{ulem}

\usepackage{amssymb,amsfonts,amsmath}
\usepackage{bm}

\usepackage[T1]{fontenc}
\usepackage[latin1]{inputenc}

\numberwithin{equation}{section}

\numberwithin{equation}{section}

\begin{document}

\title{Benchmarking Compressed Sensing, Super-Resolution, and Filter Diagonalization}

\author{Thomas Markovich}
\affiliation{Department of Chemistry and Chemical Biology, Harvard
University, Cambridge, MA 02138, USA}
\author{Samuel M. Blau}
\affiliation{Department of Chemistry and Chemical Biology, Harvard
University, Cambridge, MA 02138, USA}
\author{Jacob N. Sanders}
\affiliation{Department of Chemistry and Chemical Biology, Harvard
University, Cambridge, MA 02138, USA}
\author{Al\'{a}n Aspuru-Guzik}
\thanks{Corresponding Author. E-mail: aspuru@chemistry.harvard.edu}
\affiliation{Department of Chemistry and Chemical Biology, Harvard
University, Cambridge, MA 02138, USA}

\begin{abstract}

Signal processing techniques have been developed that use different strategies to bypass the Nyquist sampling theorem in order to recover more information than a traditional discrete Fourier transform. Here we examine three such methods: filter diagonalization, compressed sensing, and super-resolution. We apply them to a broad range of signal forms commonly found in science and engineering in order to discover when and how each method can be used most profitably. We find that filter diagonalization provides the best results for Lorentzian signals, while compressed sensing and super-resolution perform better for arbitrary signals.

\end{abstract}


\maketitle

\section{Introduction}

The reconstruction of frequency-resolved spectra from limited and undersampled measurements in the time domain is a significant problem throughout the physical sciences.  The standard approach to solving such a problem is the discrete Fourier transform, which decomposes a time series in terms of its component frequencies (or, more generally, decomposes a series into its conjugate domain).  The discrete Fourier transform offers two major advantages: no \textit{a priori} knowledge about the signal is required, and the computation can be implemented very efficiently via the fast Fourier transform.  Unfortunately, a major disadvantage is that the discrete Fourier transform imposes a natural bound on the maximum frequency resolution possible given the nature of the time series, known as the Shannon-Nyquist condition~\cite{Mallat:2008vn}.

A natural question to ask is whether the Shannon-Nyquist condition can be bypassed by exploiting any additional knowledge we may have about the signal.  Recent advances in signal processing have provided many such techniques for leveraging additional \textit{a priori} knowledge about the signal to improve reconstruction.  Our goal in this paper is to compare three such methods, filter diagonalization~\cite{1998JMagR.134...76H,4959603,Chen:1999wc,Wall:1995ur,Mandelshtam:2001vl,Mandelshtam:1997up}, compressed sensing~\cite{Donoho:ci,5272200,5288845,Sanders:2012tk,4770164,Kekatos:2011vd,4959603,Andrade28082012}, and super-resolution~\cite{5193030,Mccutchen:1967uf,2005A&A...436..373P,Freeman:2002va,CPA:CPA21455,Mallat:2008vn}, against a series of test signals in order to understand their relative strengths and weaknesses. Our comparison will be based on a subset of the signals contained in the Sparco toolbox~\cite{VanDenBerg:2007vz}, a Gaussian, a sum of random Lorentzians, and the Jacob's Ladder signal~\cite{1998JMagR.134...76H}. The Sparco toolbox provides a standard set of signal processing benchmarks while the other signals are commonly encountered throughout the physical sciences.

Filter diagonalization, one of the earliest techniques for bypassing the Shannon-Nyquist condition, assumes that the time series is generated by an underlying dynamical system with a frequency spectrum modeled by a sum of Lorentzians.  It attempts to express the frequency spectrum as a sum of Lorentzian peaks by finding the optimal frequencies, linewidths, and intensities that fit the time series. Filter diagonalization has been applied to a broad range of signals that vary from NMR spectra~\cite{1998JMagR.134...76H,Mandelshtam:2001vl} to scattering data~\cite{Wall:1995ur} and image analysis~\cite{Chen:1999wc}.

More recently, $\mathcal{L}_1$ minimization techniques, such as compressed sensing and super-resolution, have also been proposed as an alternative technique for sampling below the rate imposed by the Nyquist-Shannon condition.  Rather than assuming a particular kind of underlying dynamical system, these techniques simply assume that the signal is sparse in some \textit{a priori} known basis.  The two methods differ in both sampling strategy and the particular optimization problem to be solved.  Compressed sensing is designed to recover sparse frequency spectra (or other signals) by randomly undersampling data over the entire time domain, and then minimizing the $\mathcal{L}_1$ norm of an underdetermined system of linear equations.  Compressed sensing has been successfully applied to data acquisition in many different areas~\cite{Baraniuk:2010tc}, including the improvement of the resolution of medical magnetic-resonance imaging~\cite{Lustig:2007wf} and the experimental study of atomic and quantum systems~\cite{Andrade28082012,Sanders:2012tk,Shabani:2011we}.

Super-resolution is a related technique that shares the spirit of compressed sensing, but with a different sampling technique~\cite{5272200,Donoho:ci,5288845,4770164,4959603,Lustig:2007wf,6426647,Tuma:2009gb,5419072,4472247,5495209}. Super-resolution was developed to recover sparse frequency spectra (or other signals) from regularly sampled data over a short segment of the time domain.  It provides a provably convergent algorithm for the reconstruction of signals from these limited time-domain measurements by using a total-variation minimization procedure.  Like compressed sensing, super-resolution has been applied to a broad range of scientific problems, including image~\cite{Freeman:2002va} and video compression~\cite{Patti:1997vq}, image denoising~\cite{Elad:1997un}, atomistic modeling of open quantum systems~\cite{Markovich:2013vz}, astronomy~\cite{2005A&A...436..373P}, microscopy~\cite{Mccutchen:1967uf}, and medical imaging~\cite{5193030}.

The goal of this paper is to elucidate the strengths and weaknesses of the aforementioned signal processing techniques to provide a clear and coherent aid in choosing a method. To achieve this, we will first introduce the theory that underlies each method and outline our procedure for benchmarking the methods.  Then, we will introduce the test signals and compare the performance of each method on each signal.  Finally, we will present some general conclusions.

\section{Theory}
\subsection{Discrete Fourier Transform}
The Fourier transform is a cornerstone method in signal processing, as it provides a technique for decomposing an arbitrary function of time into its component frequencies:
\begin{eqnarray}
\hat{f}(\omega) = \frac{1}{\sqrt{2 \pi}} \int^{\infty}_{-\infty} e^{i \omega t} f(t) \, dt.
\end{eqnarray}
When treating a problem numerically, we often only have access to the values of the signal $f(t)$ on an equally-spaced $N$-point grid.  Accordingly, we discretize the  continuous Fourier transform to obtain the discrete Fourier transform:
\begin{eqnarray}
\hat{f}(\omega_i) = \frac{1}{\sqrt{N}} \sum_j f(t_j) e^{i \omega_i t_j}.
\end{eqnarray}
which can be reformulated as a matrix multiplication according to:
\begin{equation}
\hat{f}_i = \sum_j \mathcal{F}_{ij} f_j,
\end{equation}
where $f_j \equiv f(t_j)$, $\hat{f}_i \equiv \hat{f}(\omega_i)$, and $\mathcal{F}_{ij}$ is the Fourier operator. Here we have assumed a uniform frequency grid with a spacing of $f_s / N$, a time sampling rate of $\Delta t$, the maximum frequency that can be sampled is $1 / \Delta t$, $N$ is the number of time points, and $T$ is the time length of the signal. The Nyquist-Shannon sampling theorem states that if a function is band limited with maximum frequency $\Omega$, it is completely characterized with a uniform series of time points spaced by $1 / 2 \Omega$. It is often more convenient to use the converse statement in signal reconstruction, which claims that the with a sampling rate of $\Delta t$ the maximum frequency that can be recovered is $1 / 2 \Delta t$. This is a direct consequence of discretizing the Fourier transform. A major disadvantage of the discrete Fourier transform is that a long and uniformly-sampled time series is required to obtain good resolution in the frequency domain~\cite{Candes:eq, Mallat:2008vn}.

\subsection{$\mathcal{L}_1$ Minimization}
$\mathcal{L}_1$ minimization methods, including compressed sensing and super-resolution, have emerged as a powerful technique for bypassing the constraint of the Shannon-Nyquist theorem in the special case where the signal is known to be sparse in a particular basis\cite{CPA:CPA21455,Donoho:ci}.

To illustrate this, suppose we have an unknown function $f(t)$ that we wish to recover with as few samples as possible.  Suppose further that we can find a set of basis functions $\{g_i(t)\}$ such that $f(t)$ is sparse when expanded in this basis. That is,
\begin{equation}
f(t) = \sum_j \lambda_j g_j(t),
\end{equation}
where most of the $\lambda_j$ expansion coefficients are equal to zero (or near zero).  Our goal is to find the set of coefficients $\{\lambda_j\}$, since this would in turn identify the function $f(t)$.  All we know \textit{a priori} is that most of the $\lambda_j$ are zero; we do not know which of them are zero, and we do not know their values in general.  By sampling $f(t)$ at a set of points $\{t_i\}$, we can obtain a set of linear equations,
\begin{equation}\label{eq:inversion}
f_i = \sum_j  \lambda_j g_{ij},
\end{equation}
where $f_i \equiv f(t_i)$ and $g_{ij} \equiv g_j(t_i)$, and our goal is to solve these equations for the set of coefficients $\{\lambda_j\}$. While the $t$ variable suggests discretization in time, we are free to sample the signal in any domain.  Since we are trying to obtain accurate resolution by taking as few time samples as possible, in general this system of equations will be underdetermined and we must impose additional constraints to pick out the desired solution.

In $\mathcal{L}_1$ optimization methods, including compressed sensing and super-resolution, the desired solution to the underdetermined system of equations~\ref{eq:inversion} is chosen by solving the following $\mathcal{L}_1$ minimization problem:
\begin{equation}\label{eq:CS}
\begin{aligned}
 \underset{\lambda_j}{\text{argmin}} \;\; ||\lambda_j ||_1 \;\; \textrm{subject to} \;\;\;\;  || f_i - \sum_j \lambda_j g_{ij} ||_2 < \eta,
\end{aligned}
\end{equation}
where $\eta$ is a small thresholding parameter.  In this minimization problem, the $\mathcal{L}_1$ norm serves as a proxy for the sparsity-enforcing $\mathcal{L}_0$ norm by selecting the sparsest set of coefficients $\{\lambda_j\}$ such that the system of equations~\ref{eq:inversion} are satisfied to within $\eta$.  It is important to note that each $g_j(t)$ should be normalized to unity so that no single basis function is privileged.

Compressed sensing and super-resolution differ in the sampling strategy, which, in turn, is often determined by computational and experimental constraints.  Compressed sensing addresses the case where the value of the function $f(t)$ is sampled at random points $\{t_i\}$ over the entire domain.  This random sampling of points $f_i$ ensures that each point provides the maximum possible amount of information for the reconstruction of the signal.  A key result from compressed sensing is that the number of time samples $f_i$ which must be measured for accurate recovery scales roughly with the sparsity of the basis expansion (i.e. the number of nonzero $\lambda_j$), rather than the total size of the basis expansion (i.e. the total number of $\lambda_j$)~\cite{Candes:eq,Donoho:ci}.

A related method to compressed sensing is super-resolution.  Unlike compressed sensing, which applies to randomly-sampled data, super-resolution applies to data that is regularly sampled on a short segment of the time domain.  
It has been proven that super-resolution enables the recovery of signals with frequencies at one quarter of the Shannon-Nyquist condition reliably~\cite{CPA:CPA21455}.

A major advantage of both compressed sensing and super-resolution is that we can recover $f(t)$ in any basis in which the signal is sparse.  The methods work with bases as varied as wavelets~\cite{Plonka:2011vd,Duarte:2008wx}, treelets~\cite{Lee:2008ve}, geometric harmonics~\cite{Coifman:2005uy}, and polynomials~\cite{Kekatos:2011vd,Li:2010ve}.  All that is required is that we know the sparse basis ahead of time.  Although this may seem like a strong restriction, for many scientific problems physical intuition often leads to a sparse basis.  One does not need to pick the optimal basis; any reasonably sparse basis will work.  Moreover, both compressed sensing and super-resolution are robust to choosing an overcomplete basis, which allows for a lot more freedom in finding a sparse basis.

For example, in computational chemistry, we are often interested in resolving spectra which are known to be sparse directly in the frequency domain (i.e. the spectrum is mostly zero except for a few sharp frequency peaks).  In this case, we might choose a basis of complex exponentials $g_j(t) = \frac{1}{2\pi} e^{i \omega_j t}$.  After time sampling, the matrix $g_{ij} = \frac{1}{2\pi}e^{i \omega_j t_i}$ simply becomes an undersampled set of rows of the discrete Fourier transform matrix.  Once the sparse coefficients $\lambda_j$ have been found by solving eq.~\ref{eq:CS}, the final spectrum may be plotted as
\begin{equation}
\hat{f}(\omega) = \sum_j \lambda_j \hat{g}_j(\omega) = \sum_j \lambda_j \delta \left( \omega - \omega_j \right),
\end{equation}
Other similar bases commonly used when applying compressed sensing or super-resolution to Fourier analysis are sine functions $g_j(t) = \frac{1}{2\pi} \sin \left( \omega_j t \right)$ and cosine functions $g_j(t) = \frac{1}{2\pi} \cos \left( \omega_j t \right)$.

To take another common example in the physical sciences, we often find damped oscillatory signals which may be expressed as a sum of damped cosines:
\begin{equation}
g_{jk}(t) = e^{-\gamma_k \, t} \cos \left( \omega_j t \right).
\end{equation}
Compressed sensing and super-resolution are easily adapted to this overcomplete basis and, once the sparse coefficients $\lambda_{jk}$ have been found via eq.~\ref{eq:CS}, the final spectrum may be plotted as
\begin{eqnarray}
\hat{f}(\omega) &=& \sum_{j,k} \lambda_{jk} \hat{g}_{jk}(\omega) \\
&=& \sum_{j,k} \dfrac{\lambda_{jk}}{\sqrt{2 \pi}} \left( \frac{\gamma_k }{\gamma_k^2 + \left( \omega - \omega_j \right)^2 } + \frac{ \gamma_k }{\gamma_k^2 + \left( \omega + \omega_j \right)^2 }  \right). \nonumber
\end{eqnarray}

In short, compressed sensing and super-resolution both enable the recovery of an undersampled signal by using a customized, sparse basis that is appropriate to the problem at hand.  The choice of technique typically depends on which sampling method is easier to perform: random sampling over the entire time domain is appropriate for compressed sensing, while regular sampling over a short part of the time domain is appropriate for super-resolution.

\subsection{Filter Diagonalization}
Filter diagonalization is another approach to circumvent the Shannon-Nyquist condition. Inspired by quantum mechanics, the method assumes that the signal $f(t)$ to be recovered is generated by the time evolution of a unitary propagator,
\begin{equation}\label{fdm:correlation}
f(t) = ( \Phi_0 , e^{-i \hat{\Omega} t}  \Phi_0 ).
\end{equation}
If we sample $f(t)$ on an equally-spaced grid $t_n = n \tau$, we can discretize this equation as
\begin{equation}\label{fdm:correlation2}
f(t_n) = ( \Phi_0 ,  e^{-i n \hat{\Omega} \tau} \Phi_0 ).
\end{equation}
where $\hat{U} = e^{-i \hat{\Omega} \tau}$ is the unitary propagator.  By expanding the propagator in terms of its eigenvalues and (possibly complex) eigenvectors,
\begin{equation}\label{fdm:expansion}
e^{-i \hat{\Omega} \tau} = \sum_j e^{-i \omega_j \tau} |u_j ) ( u_j |,
\end{equation}
and substituting this expansion into \ref{fdm:correlation2}, we obtain
\begin{equation}\label{fdm:lorentzian}
f(t_n) = \sum_j | ( u_j , \Phi_0 ) |^2 e^{-i \omega_j n \tau},
\end{equation}
which is the equation for a Lorentzian signal with (possibly damped) frequencies $\omega_j$ and amplitudes $\lambda_j \equiv | ( u_j , \Phi_0 ) |^2$.  Hence, resolving a signal $f(t)$ into a sum of Lorentzian peaks is reduced to the standard linear algebra problem of finding the eigenvalues and eigenvectors of the propagator $\hat{U} = e^{-i \hat{\Omega} \tau}$.

The key insight of filter diagonalization is that the propagator to be diagonalized, $e^{-i \hat{\Omega} \tau}$, may be expressed entirely in terms of time samples of the signal $f_n \equiv f(t_n)$.  A common approach is to write the propagator in a so-called Krylov basis,
\begin{equation}\label{fdm:krylov}
\Psi_k = \sum^N_{n=0} \left(\frac{\hat{U}}{z_k}\right)^n \Phi_0,
\end{equation}
where $z_k = e^{-i \nu_k \tau}$ is a complex value chosen along the unit circle.  By selecting the $\nu_k$ close to the frequencies we wish to resolve, it is possible to filter $f(t)$ and recover only those frequency components near the $\nu_k$; this is where the name filter diagonalization comes from.  It is important to include more basis vectors $|\Psi_k )$ than there are frequencies we wish to resolve.

Expressing the propagator in the Krylov basis yields
\begin{equation}\label{fdm:propagator}
U_{kk'} = ( \Psi_k , \hat{U}  \Psi_{k'} ) = \sum_{n=0}^N \sum_{n' = 0}^N f_{n+n'+1} z_k^{-n} z_{k'}^{-n'},
\end{equation}
which is expressed completely in terms of time samples of the signal.  Because the Krylov basis is not orthonormal, we also need the overlap matrix
\begin{equation}\label{fdm:overlap}
S_{kk'} = ( \Psi_k , \Psi_{k'} ) = \sum_{n=0}^N \sum_{n' = 0}^N f_{n+n'} z_k^{-n} z_{k'}^{-n'},
\end{equation}
after which the eigenvalues and eigenvectors of $\hat{U}$ may be found by solving the generalized eigenvalue problem,
\begin{equation}\label{fdm:eigenvalue}
U B_j = u_j S B_j.
\end{equation}
For computational efficiency, the double sums in eqs.~\ref{fdm:propagator} and \ref{fdm:overlap} are rewritten as single sums, as shown in~\cite{Chen:1999wc}. With the eigenvalues $u_j$ and eigenvectors
\begin{equation}\label{fdm:eigenvectors}
u_j  = \sum_k B_{kj} \Psi_k
\end{equation}
in hand, the frequencies $\omega_j$ and amplitudes $\lambda_j$ in the signal $f(t)$ may be reconstructed according to the formulas
\begin{align}\label{fdm:reconstruction}
u_j &= e^{-i \omega_j \tau},\textrm{ and}\\
\lambda_j &= \left| ( u_j , \Phi_0 ) \right|^2 = \left| \sum_k B_{kj} ( \Psi_k , \Phi_0 ) \right|^2.
\end{align}

Because the Krylov basis is often close to becoming linearly dependent, we include a numerical conditioning step to remove possible spurious frequencies.  In particular, we select a value of $p$ and resolve the generalized eigenvalue equation with $\hat{U}^{p+1}$ and $\hat{U}^{p}$ (used in place of $\hat{U}$ and $\hat{S}$).  We remove the eigenvalues that are not shared in the two spectra, and then select a filtering grid with frequencies $\nu_j$ located only at the nonspurious eigenvalues.  We rerun filter diagonalization one more time on this adaptive frequency grid~\cite{Mandelshtam:1997up}, and these are the results we report below.

While this technique was initially derived with quantum mechanics in mind, it is not limited to such applications. Indeed, with generalizations such at 2D filter diagonalization and multi-resolution filter diagonalization, the method has been expanded to be applicable to a broad range of signals.

\section{Methods}
As our goal in this paper is to compare the performance of compressed sensing, super-resolution, and filter diagonalization in recovering sparse signals, we began by obtaining a series of sparse signals from the Sparco toolkit, which is a well-known set of sparse signals used for benchmarking various signal processing techniques~\cite{VanDenBerg:2007vz}. We also generated a few other signals of interest to highlight particular properties of each technique. Unless otherwise stated, each signal began as a continuous function of time $f(t)$ and, to generate a discrete time series, we sampled $f(t)$ at 4096 time points ranging uniformly from $t = 0$ to $t = 1$ second. This gave a grid separation of $1/4096$ seconds, with a maximum recoverable frequency of $2048$ Hz.

For each sparse signal processing method, we varied how many of the 4096 time points we sampled (in increments of 64) and investigated the dependence of the recovery error on the extent of undersampling.  As a measure of the recovery error, we employed the relative 2-norm error over all 4096 time points (regardless of the extent of undersampling):
\begin{equation}\label{eq:error}
\textrm{Recovery Error} = \frac{\sum_{i = 1}^{4096} |f_{\textrm{recovered}}(t_i) - f_{\textrm{original}}(t_i)|^2}{\sum_{i = 1}^{4096} | f_{\textrm{original}}(t_i)|^2}
\end{equation}
We consistently obtained similar results with the 1-norm error and the $\infty$-norm error, but the 2-norm error has the advantage that, by Parseval's theorem, it is the same whether it is measured in the time domain or the frequency domain. Therefore, we adopted the 2-norm as our primary benchmark. In some instances, filter diagonalization has been marketed as a parameter estimation technique, but this problem is identical to reconstruction of the signal as a whole. That is, if a method can extract the characteristic parameters of the signal, then these parameters can be used to reconstruct the signal.

For compressed sensing and super-resolution, we attempted to recover each signal in an appropriate sparse basis. The basis used depends on the signal and is discussed in the individual sections below. Because super-resolution
requires a grid of equally spaced sample points, we began our analysis by examining the full signal and computing the errors. We then repeated our analysis for the signal by successively undersampling in powers of two, taking care to ensure that our sample points were always equally spaced. In contrast, our analysis with compressed sensing involved randomly selecting the same number of points that were included in the super-resolution analysis at each step.

For filter diagonalization, we used the same regular sampling strategy as for super-resolution, and we monitored the recovery error as a function of the sampling.  Filter diagonalization requires specifying a grid of frequencies on which we expect the components of the signal to lie, so we specified a frequency range of 0 kHz to 20 kHz. To find the appropriate grid density and number of frequencies, we tuned these two parameters for optimal reconstruction with the full time signal and assumed these parameters would be valid for the entire numerical experiment.  To ensure the robustness of our results, we also performed the analysis from 0 to 5, 10, 50, 100, 200, and 500 kHz, with a similar density of frequencies.

From a numerics standpoint, compressed sensing and super-resolution require a fast, memory-efficient $\mathcal{L}_1$ solver.  For all results in this paper, we implemented the two step iterative shrinkage/thresholding (TwIST) algorithm in Python~\cite{BioucasDias:2007tp,4378902}.  Our TwIST solver is capable of solving arbitrary optimization problems given a measurement matrix, signal vector, and objective function, and gives numerically identical answers to the Matlab version for a wide range of test signals, including all those in this paper.

We implemented filter diagonalization in Python, performing all required matrix diagonalizations using the zgeev function from LAPACK.  We benchmarked our implementation against Harminv, a freely-available C++ implementation based on the methods described in~\cite{Mandelshtam:1997up}, and found that they give the same answers to within numerical precision for a wide range of signals, including all of those presented in this paper.

\subsection{Gaussian Signal}

\begin{figure}
        \centering
                \begin{subfigure}[b]{0.4\textwidth}
                \includegraphics[width=1.1\textwidth]{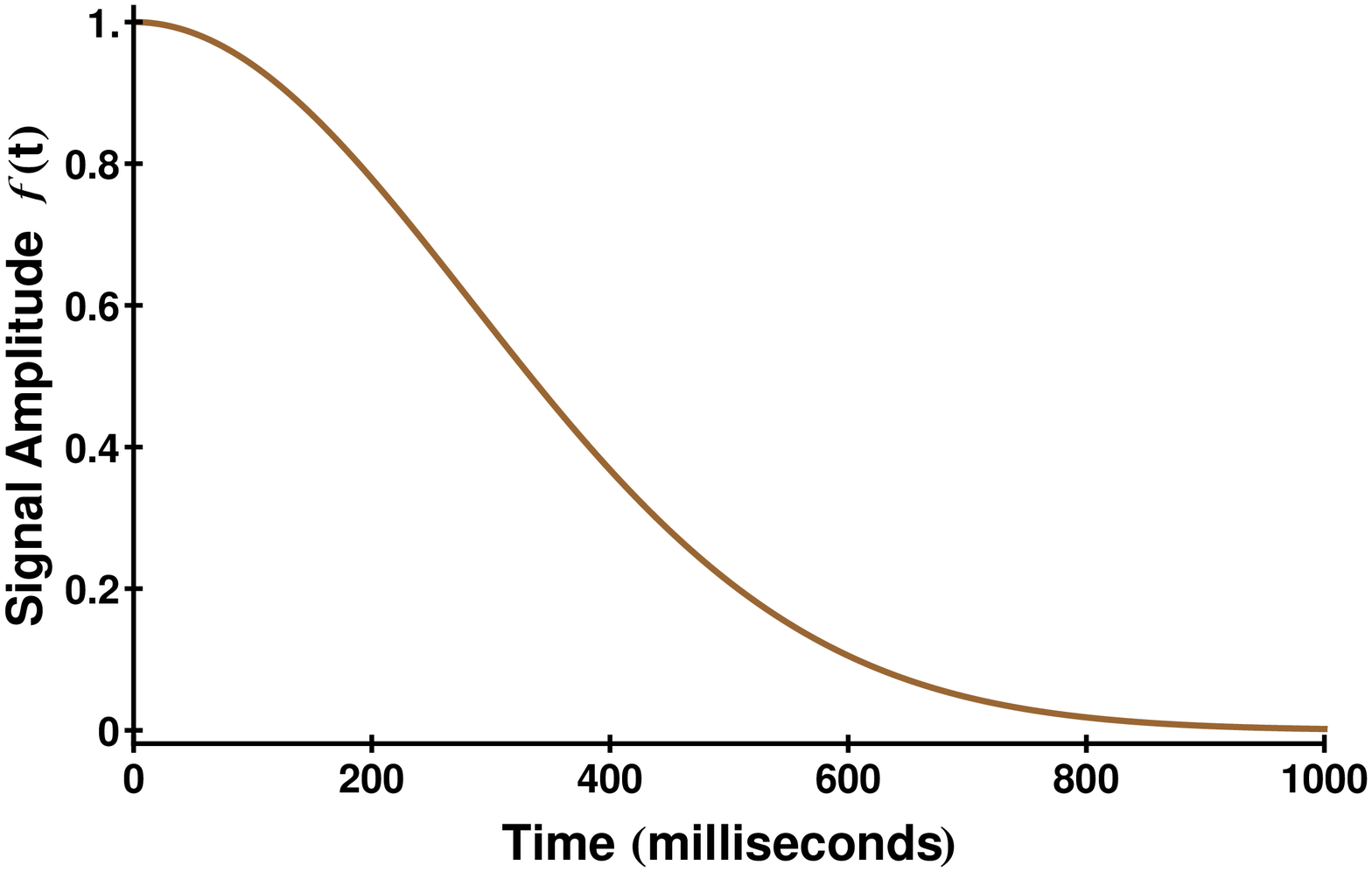}
                \caption{}
                \label{fig:gaussian}
        \end{subfigure}
                        \qquad
                \begin{subfigure}[b]{0.4\textwidth}
                \includegraphics[width=1.1\textwidth]{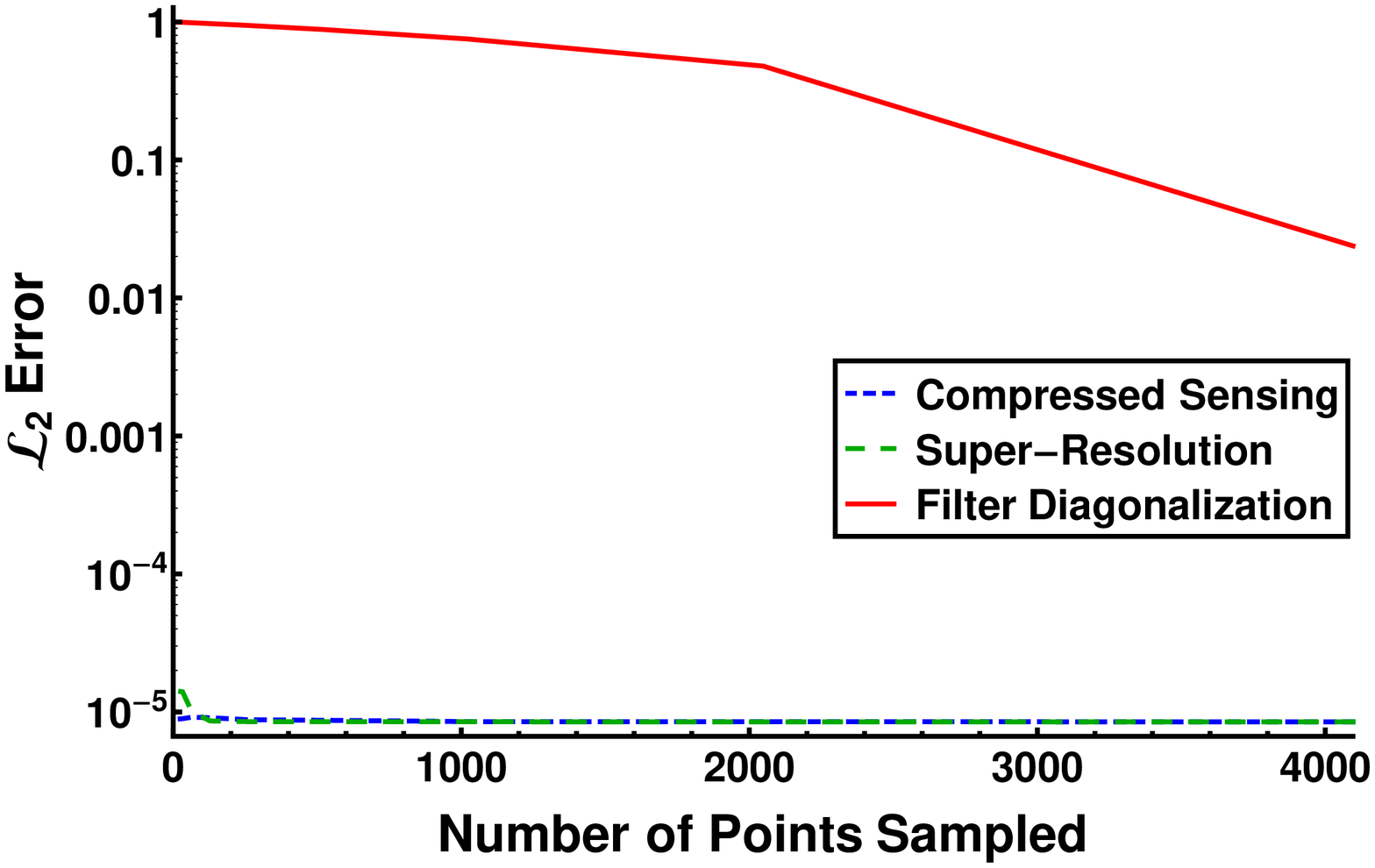}
                \caption{}
                \label{fig:gaussianErrors}
        \end{subfigure}
        \caption{(a) Time series consisting of a Gaussian given by Equation \eqref{eq:gaussian} centered at $t = 0$ with standard deviation $\sigma = 0.4$.  (b) Comparison of the relative 2-norm error in the reproduction of a Gaussian signal as a function of undersampling betweening compressed sensing, super-resolution, and filter diagonalization.}
\end{figure}

We begin with one of the most ubiquitous signals throughout signal processing, a simple Gaussian centered at time $t = 0$ with $\sigma = 0.4$ (Fig.~\ref{fig:gaussian}),
\begin{equation}\label{eq:gaussian}
f(t) = e^{-\frac{t^2}{0.4^2}}.
\end{equation}

To recover this signal with compressed sensing and super-resolution, we employ a basis of displaced Gaussians
\begin{equation}
g_{jk}(t) = e^{-\frac{(t-t_j)^2}{\sigma_k^2}},
\end{equation}
with 100 centers $t_j$ ranging uniformly from 0 to 1, and 100 standard deviations $\sigma_k$ also ranging uniformly from 0 to 1, for a total of 10,000 different basis functions.  It is clear that the function we hope to recover, $f(t)$, is sparse in this basis.

Fig.~\ref{fig:gaussianErrors} compares the performance of compressed sensing, super-resolution, and filter diagonalization in recovering the Gaussian signal.  Compressed sensing and super-resolution both converge quickly to the correct signal, and as more time-domain information is sampled, the signal becomes more obviously composed of a single Gaussian.  Moreover, compressed sensing converges more quickly than super-resolution, indicating that randomly sampling over the entire time domain provides more complete information about the overall shape of the signal than sampling uniformly with a coarse grid. Both compressed sensing and super-resolution recover a single strongly converged, correct, peak with amplitude 1 and a few spurious peaks with amplitudes smaller than $10^{-6}$. This represents a small numerical instability in our implementation of TwIST but these spurious features are easy to identify and disregard.

By contrast, filter diagonalization fails to converge completely because it attempts to recover the Gaussian as a sum of Lorentzian peaks, rather than taking advantage of the natural sparsity of the signal in a Gaussian basis. This example highlights the basis set agnosticism of the $\mathcal{L}_1$ minimization techniques, which is one of their principal advantages.

\subsection{Sparco Problem 1}
\begin{figure}
        \centering
        \begin{subfigure}[b]{0.4\textwidth}
                \includegraphics[width=1.1\textwidth]{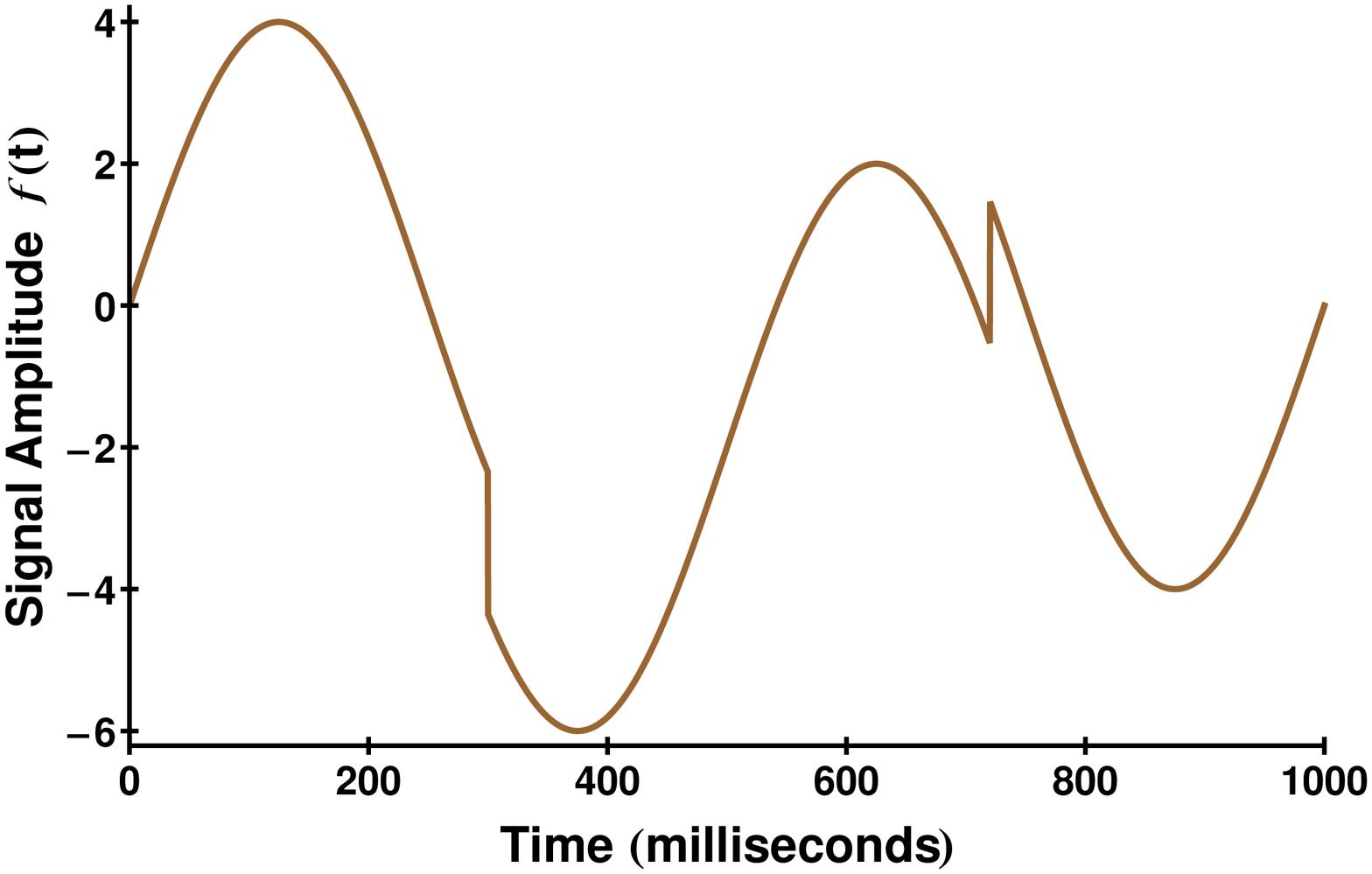}
                \caption{}
                \label{fig:signal001}
        \end{subfigure}
        \qquad
        \begin{subfigure}[b]{0.4\textwidth}
                \includegraphics[width=1.1\textwidth]{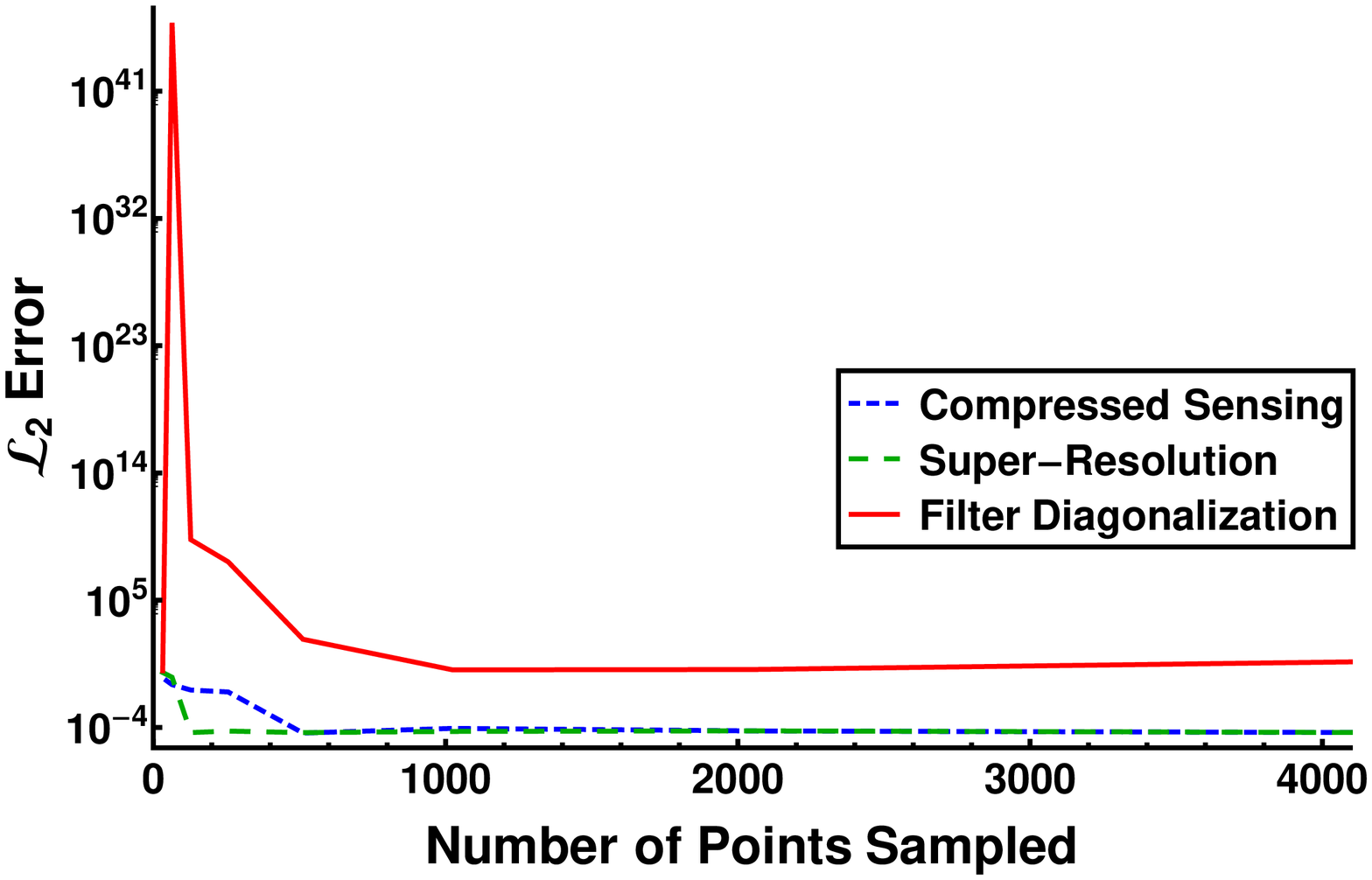}
                \caption{}
                \label{fig:signal001Errors}
        \end{subfigure}
        \caption{(a) The combination of a sinusoid and a Heaviside signal in the time domain as given by Equation \eqref{eq:sparco1}.  (b) Comparison of the 2-norm error in the reproduction of Problem 1 from the Sparco toolbox as a function of undersampling between compressed sensing, super-resolution, and filter diagonalization.  We attribute the large spike in error by filter diagonalization to the recovery of spurious exponentially divergent solutions.}
\end{figure}

For our second signal, we consider a sinusoid that is ``disrupted'' by two Heaviside step functions (Fig.~\ref{fig:signal001}),
\begin{equation}\label{eq:sparco1}
f(t) = 4 \sin(4 \pi t) - \Theta(t - 0.3) - \Theta(0.72 - t),
\end{equation}
one of the earliest signals used to benchmark wavelet and compressed sensing techniques~\cite{Buckheit:jt, 2001SIAMR..43..129C, Donoho:ci}.  This signal is Problem 1 in the Sparco toolbox of sparse signals~\cite{VanDenBerg:2007vz}.  

To recover this signal via compressed sensing and super-resolution, we employ a composite basis of sine functions and Heaviside step functions
\begin{align*}
g_j(t) &= \sin \left( \omega_j t \right)\\
h_j(t) &= \Theta(t - t_j),
\end{align*}
with the spectral spacing $\omega_j$ of the sine functions ranging uniformly from 0 to 4096$\pi$ kHz in units of $\frac{\pi}{10}$ kHz, and the unit steps $t_j$ of the Heaviside step functions ranging uniformly from 0 to 1 second in units of 0.01 seconds.  While either basis $g_j(t)$ or $h_j(t)$ by itself would provide a complete basis for recovery of the signal $f(t)$ (to within numerical precision), the function $f(t)$ would not be sparse in either basis on its own.  On the other hand, there is no problem in $\mathcal{L}_1$ minimization techniques with using the combined basis, which affords the additional advantage that $f(t)$ is sparse in this combined basis.  However, when building a composite basis with different functional forms, it is important to ensure that each basis function is normalized to the same value, for which we chose unity.

Fig.~\ref{fig:signal001Errors} compares the performance of compressed sensing, super-resolution, and filter diagonalization in recovering the signal $f(t)$.  For super-resolution and filter diagonalization, both of which involve regular sampling over a coarse time domain grid, most of the initial error simply comes from the fact that the methods cannot reproduce aspects of the signal that have not been sufficiently sampled.

Super-resolution and compressed sensing are able to identify that the signal has some underlying sine structure, but initially fails to recognize the exact position of the step functions. As more samples are included in the analysis, both techniques are able to quickly converge to the exact location of the step function. This convergence leads to a very sharp phase transition characteristic of an $\mathcal{L}_1$ analysis, and represents the minimum amount of information required to exactly reproduce the full signal. This phase transition is a well known aspect of $\mathcal{L}_1$ minimization techniques, and provides a useful and valid check on convergence and accuracy.

This stands in contrast to filter diagonalization, which attempts to match the Heaviside step functions by creating a signal that contains exponentially growing components, eventually resulting in an explosion of error.  To better understand this behavior, we varied the magnitude of the Heaviside step functions, but found that the creation of an exponentially growing signal persisted even when the Heaviside step function was $0.1\%$ of the amplitude of the oscillating sine wave.

Not surprisingly, compressed sensing fares better than both super-resolution and filter diagonalization.  This is easily explained by the fact that compressed sensing randomly samples the entire domain, so it can quickly ``recognize'' all features of the signal and recover them accurately.  After roughly one sixty fourth of the signal has been sampled, the error changes only marginally, and this effect is robust across different runs of random sampling.

\subsection{Sparco Problem 5}
\begin{figure}
        \centering
        \begin{subfigure}[b]{0.4\textwidth}
                \includegraphics[width=1.1\textwidth]{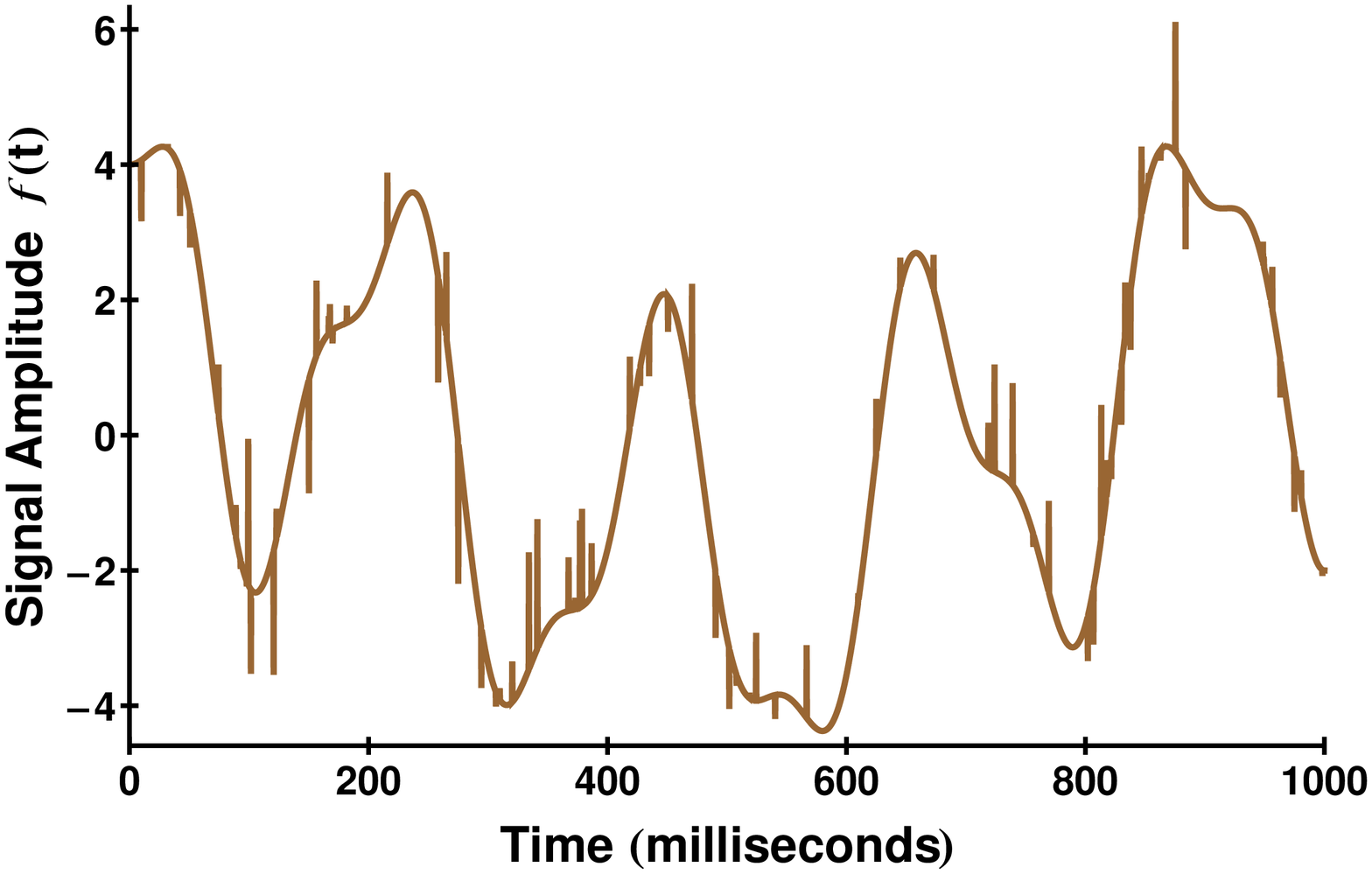}
                \caption{}
                \label{fig:signal005}
        \end{subfigure}
        \qquad
        \begin{subfigure}[b]{0.4\textwidth}
                \includegraphics[width=1.1\textwidth]{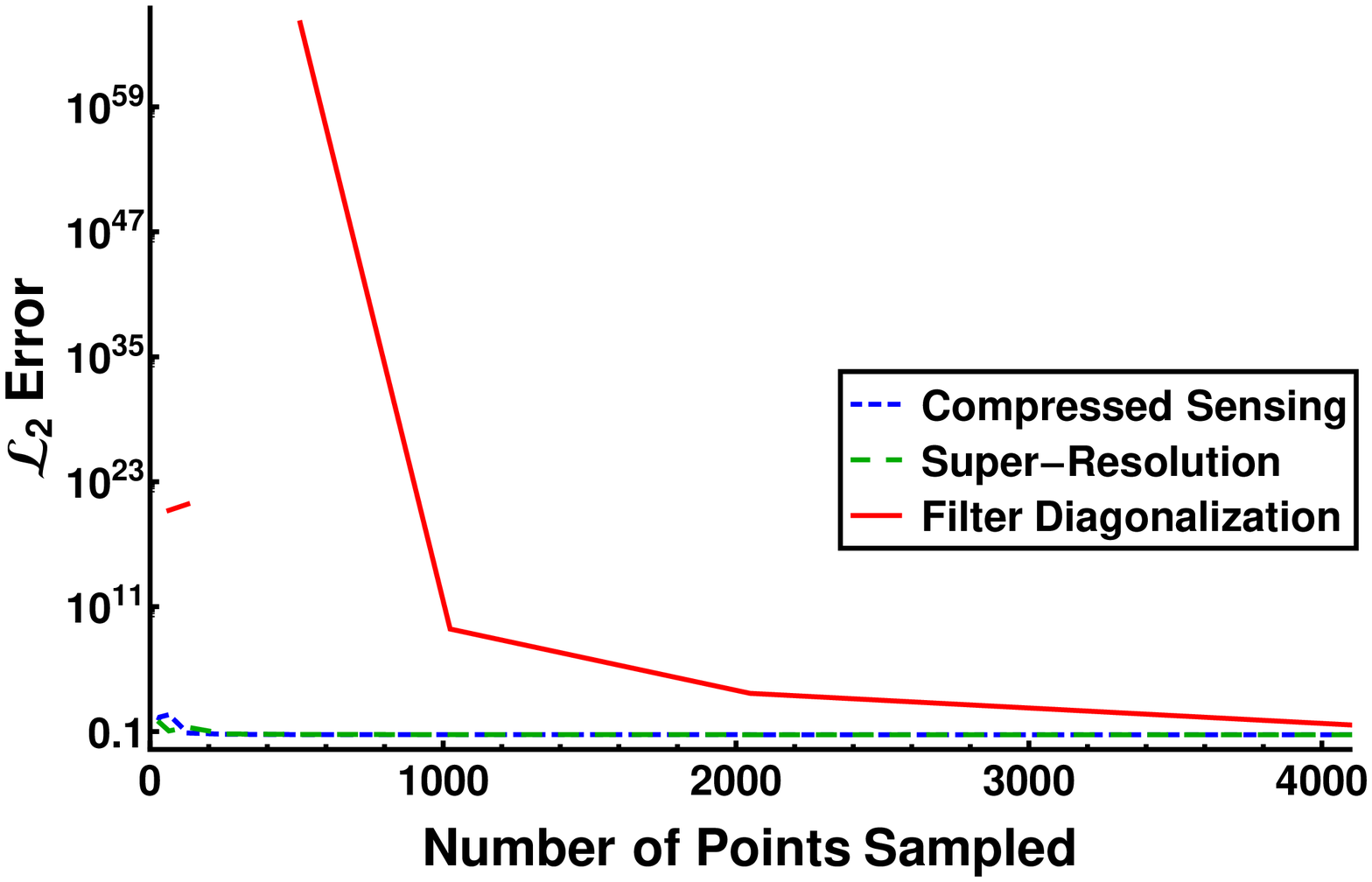}
                \caption{}
                \label{fig:signal005Errors}
        \end{subfigure}
        \caption{(a) The combination of three cosines and random spikes in the time domain as described in Equation \eqref{eq:prob5time}.  (b) Comparison of the 2-norm recovery error for the Problem 5 from the Sparco toolbox as a function of undersampling for compressed sensing, super-resolution, and filter diagonalization.  We attribute the large error peaks found by filter diagonalization to the recovery of spurious exponentially divergent solutions.}
\end{figure}

For our third signal, we consider the sum of three cosines with the addition of 40 spikes at random time points $\{t_i\}$ (Fig.~\ref{fig:signal005}),
\begin{eqnarray}\label{eq:prob5time}
f(t) &=& 2 \cos(2 \pi t) + 3 \cos(9 \pi t) - \cos(20 \pi t) \\ \nonumber
 &+& \sum_i \alpha_i \delta(t - t_i),
\end{eqnarray}
where $\delta(t-t_i)$ is regarded here as the Kronecker delta function (equal to 1 at the time point $t = t_i$, 0 otherwise) and $\alpha_i$ is a uniform random number between 0 and 1.  This signal is Problem 5 in the Sparco toolbox of sparse signals~\cite{VanDenBerg:2007vz}, and including it in our comparison is particularly useful for benchmarking the ability of compressed sensing, super-resolution, and filter diagonalization to deal with random noise.


To recover this signal via compressed sensing and super-resolution, we employ a cosine basis:
\begin{align*}
g_j(t) &= \cos \left( \omega_j t \right)
\end{align*}
with the spectral spacing $\omega_j$ of the cosine functions ranging uniformly from 0 to 4096$\pi$ kHz in units of $\frac{\pi}{10}$ kHz. Note that we do not include Kronecker delta functions $\delta(t - t_i)$ in our basis, since our goal is to see whether our signal processing methods can recover the underlying cosine functions despite the random noise.

As shown in Fig.~\ref{fig:signal005Errors}, both compressed sensing and super-resolution successfully recover the underlying cosine functions in spite of the noise peaks (the noise peaks simply get absorbed into the denoising parameter $\eta$).  As expected, compressed sensing recovers the signal with less sampling than super-resolution, since randomly sampled points over the entire time domain effectively contribute more information than regularly sampled points over that same time domain.

In contrast, filter diagonalization does not include a robust denoising procedure, and the method struggles with the Kronecker delta peaks because they represent sharp deviations from the underlying cosine signal.  In particular, in attempting to match the Kronecker delta peaks, filter diagonalization creates a signal that contains exponentially growing components rather than exponentially damped Lorentzians.

In summary, compressed sensing and super-resolution both pick out the underlying cosine signals by denoising the Kronecker delta peaks, whereas filter diagonalization does not.

\subsection{Jacob's Ladder}
\begin{figure}
        \centering
        \begin{subfigure}[b]{0.4\textwidth}
                \includegraphics[width=1.1\textwidth]{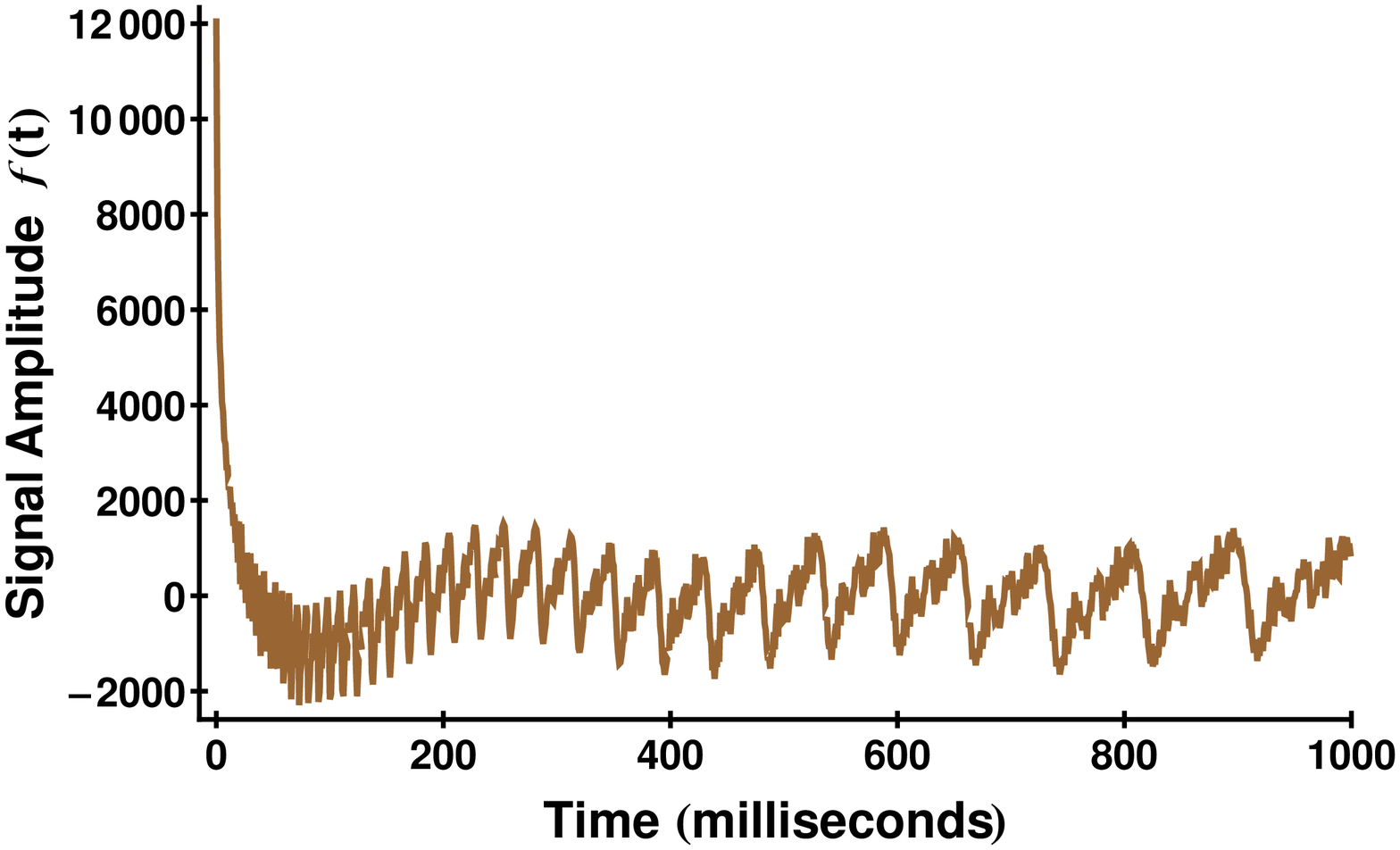}
                \caption{}
                \label{fig:jacobLadder}
        \end{subfigure}
        \qquad
        \begin{subfigure}[b]{0.4\textwidth}
                \includegraphics[width=1.1\textwidth]{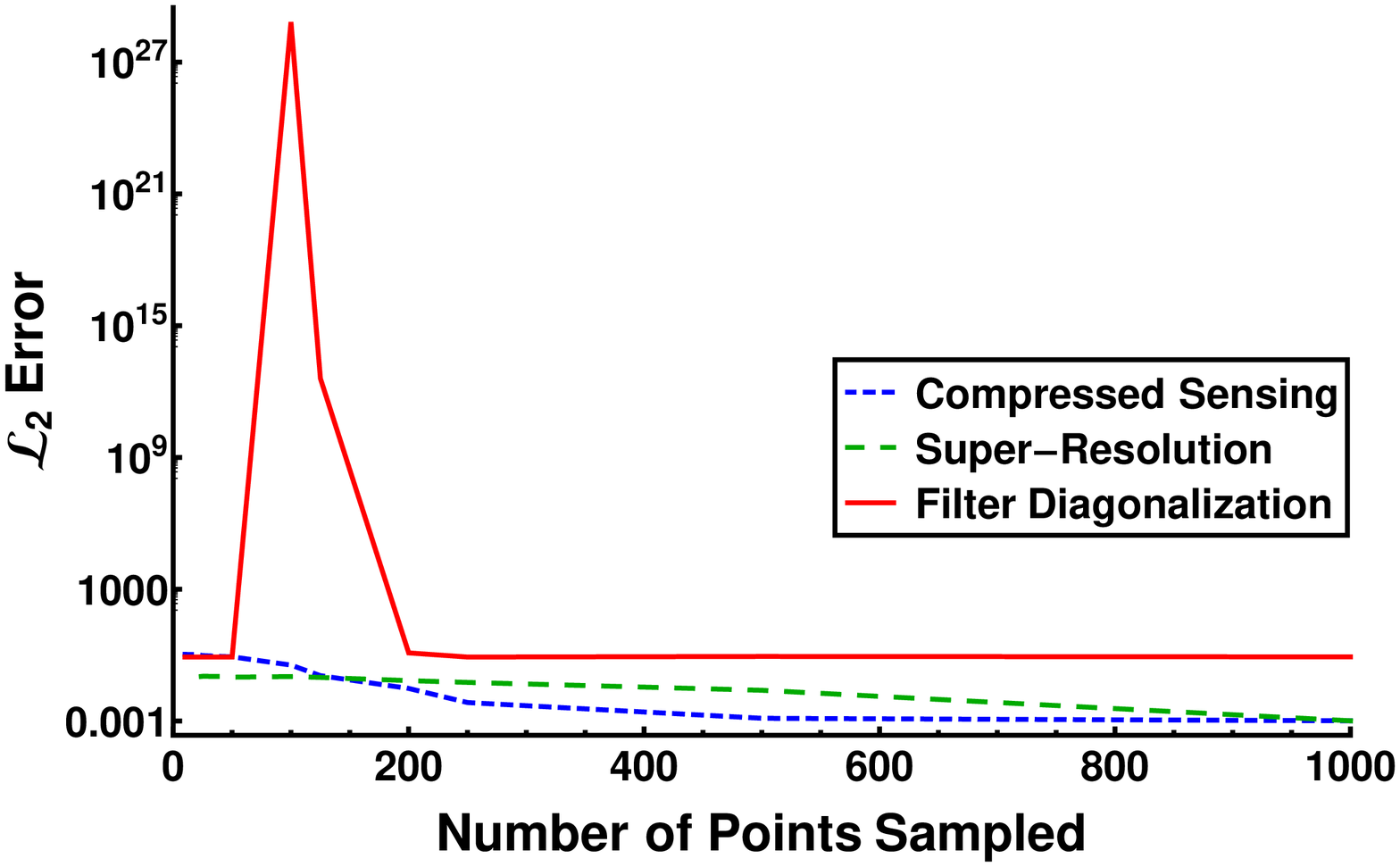}
                \caption{}
                \label{fig:jacobLadderErrors}
        \end{subfigure}
        \caption{(a) The Jacob's Ladder signal, a combination of many Lorentzian peaks, in the time domain as given by Eq. \eqref{eq:JL}.  (b) Comparison of the 2-norm error for the reproduction of the Jacob's Ladder signal as a function of undersampling for compressed sensing, super-resolution, and filter diagonalization.  We have no explanation clear for why our implementation of filter diagonalization fails to reproduce this signal, and we note that further work may be necessary.}
\end{figure}

Next, we consider a time series devised by some of the original developers of filter diagonalization for benchmarking sparse signal processing techniques~\cite{1998JMagR.134...76H}.  This signal is known as Jacob's Ladder, and it consists of a very large number of Lorentzian peaks (Fig.~\ref{fig:jacobLadder}):

\begin{eqnarray}
\label{eq:JL}
f_n &=& \sum_{m=0}^{49} e^{-1.8 m \pi 2\times10^{-4} * n} \left[ \cos(1.8 m \pi 2\times10^{-4}  2500 n) \right. \\
 &+& \left. \cos(1.8 m \pi 2\times10^{-4}  2487.5 n)+\cos(1.8 m \pi 2\times10^{-4}  2475.0 n) \right] \nonumber
\end{eqnarray}

In this experiment, we created a signal with 1000 data points that ranged, uniformly, from 0 to 1 seconds using the formula in Equation \eqref{eq:JL}. We first analyzed this signal with a numerical implementation of filter diagonalization using a frequency range of 0 kHz to $\pi$ kHz and assumed that we had a maximum of 4500 frequencies in our signal.

We started by performing the filter diagonalization analysis with the above frequency grid using the entire signal, and used the recovered frequencies and expansion coefficients to construct the recovered signal. This allowed us to compute the 2-norm error between the recovered signal and the exact signal over the entire range. We then repeated the analysis, while successively undersampling, first taking every second point, then every fourth, fifth, eighth, tenth, twentieth, twenty-fifth, fortieth, fiftieth, and finally, hundredth. We then performed the same analysis with super-resolution and compressed sensing. Because compressed sensing involved random sampling, we took care to randomly sample the same number of points that were included in the super-resolution analysis at each step.

In order to perform the $\mathcal{L}_1$ analysis we examined the functional form and the signal itself and concluded that it should be sparse in the basis of damped oscillators. Thus we constructed such a basis with a spectral spacing of $0.01$ Hz and a maximum of 4$\pi$ Hz. We also scanned exponential decay parameters ranging from 0 to $\pi$ Hz in steps of 0.01 Hz. For the super resolution analysis, we performed the same time addition procedure that was performed with filter diagonalization. We began the compressed sensing analysis by picking 50 random time points, and at each subsequent step an additional 50 random points were taken from those remaining until we were sampling the full signal.

From the errors given in Figure~\ref{fig:jacobLadderErrors}, both compressed sensing and super resolution converge to a better answer more rapidly than filter diagonalization. We attribute these errors to the recovery of exponentially divergent solutions but we have no way of accounting for the difference between our results and those obtained elsewhere. Our working theory is a sensitivity of filter diagonalization to the frequency grid and parameter choice, and we have not found the correct combination of parameters that allows us to completely recover the desired signal. This suggests that our implementation, as well as our standard of comparison suffered from significant numerical instability. We are unsure whether this is explained better by fundamental instabilities in the method at hand, or simply instabilities in the current implementations.

By contrast, super-resolution and compressed sensing do not suffer from these same problems. Many current methods are extremely stable. These techniques only include the basis functions that are explicitly chosen. Unfortunately, we are limited to the recovery of Lorentzian parameters (frequency and line-width), that are on the grid, which requires a sufficiently dense set of parameters for accurate recovery. As a result, $\mathcal{L}^1$-optimization methods become more and more memory intensive as the number of basis functions increases.

\subsection{Sum of Random Lorentzians}

\begin{figure}
        \centering
        \begin{subfigure}[b]{0.4\textwidth}
                \includegraphics[width=1.1\textwidth]{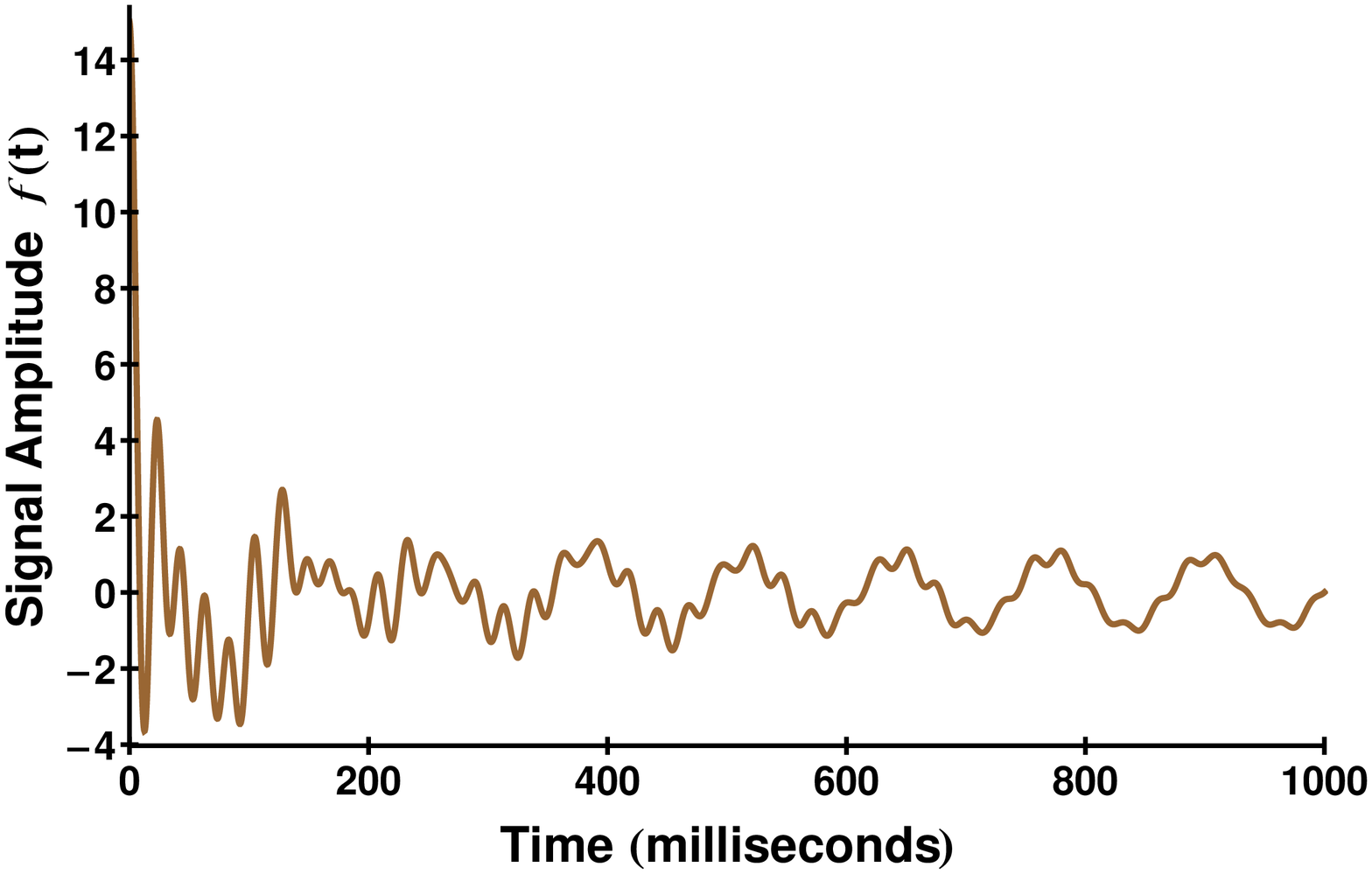}
                \caption{}
                \label{fig:sorl}
        \end{subfigure}
        \qquad
        \begin{subfigure}[b]{0.4\textwidth}
                \includegraphics[width=1.3\textwidth]{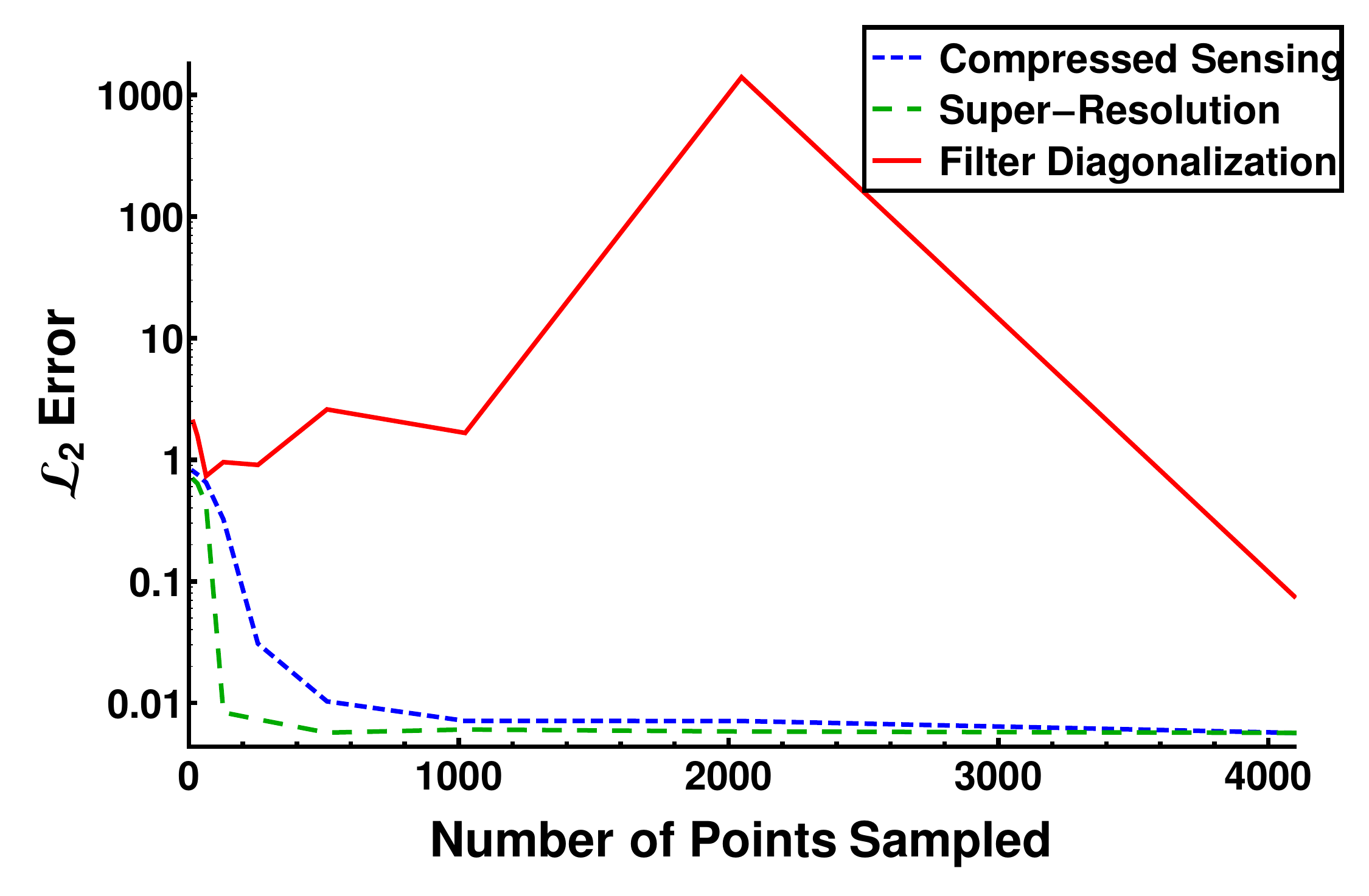}
                \caption{}
                \label{fig:sorlErrors}
        \end{subfigure}
        \caption{(a) Combination of twenty random damped cosines as described in Eq.~\eqref{eq:sorlTime}.  (b) Comparison among compressed sensing, super-resolution, and filter diagonalization of the 2-norm error in the reproduction of the combination of random damped cosines as a function of undersampling.  We attribute the large error peaks found by filter diagonalization to the recovery of spurious exponentially divergent solutions.}
\end{figure}

For the final comparison, we created a sum of twenty random damped cosines:
\begin{equation}\label{eq:sorlTime}
f(t) = \sum_{n=1}^{20} e^{-\gamma_n t} \cos{\omega_n t}
\end{equation}
with $\gamma_n$ drawn from a uniform random distribution ranging from 0 to 20 Hz and $\omega_n$ ranging from 0 to 50$\pi$ Hz. This type of autocorrelation signal is ubiquitous not only in chemistry applications, but in signal processing at large.

To recover this signal via compressed sensing and super-resolution, we employ a basis of damped cosine functions:
\begin{align*}
g_{jk}(t) &= e^{-\gamma_k t} \cos \left( \omega_j t \right).
\end{align*}
Here, the spectral spacing $\omega_j$ ranges uniformly from 0 to 40$\pi$ Hz in steps of $\pi/24$ Hz and the damping parameters $\gamma_k$ ranges uniformly from 0 to 20 Hz in steps of $1/2$ Hz.  For filter diagonalization, we selected a frequency range from 0 to 100$\pi$ Hz and chose a basis of 1200 frequencies. This was chosen because it gave near perfect reconstruction of the full signal, while bases smaller than this were prone to numerical instability.

As shown in Fig.~\ref{fig:sorlErrors}, both compressed sensing and super-resolution successfully recover the underlying damped cosine structure but is restricted to the functions on the grid. This is the most significant source of error.  As expected, compressed sensing recovers the signal with less sampling than super-resolution, since randomly sampled points over the entire time domain effectively contribute more information than regularly sampled points over a short time.

In contrast, filter diagonalization can recover off-grid frequencies extremely efficiently. Because of this, the final errors should be smaller than the errors from both compressed sensing and super resolution. Unfortunately, we encountered significant stability issues during many of our decompositions that resulted in exponentially growing solutions. These solutions gave $\mathcal{L}_2$ errors on the order of $10^{4}$ at times (note that we have chosen to only plot a few orders of magnitude in ~\ref{fig:sorlErrors}.) While filter diagonalization is capable of giving a much better answer, the technique is significantly more sensitive to slight deviations in the operational parameters chosen.

\section{Conclusions}
In conclusion, we have performed a broad comparison of three different signal processing techniques that attempt to ``beat'' the Shannon-Nyquist limit. With prior information about a reasonable basis for your signal, $\mathcal{L}_1$ minimization techniques provide a robust and faithful reproduction of the signal. We emphasize that the difference between super-resolution or compressed sensing is simply a choice of sampling procedure and normally is determined by the data acquisition technique.

Additionally, we found that if the signal at hand was Lorentzian, filter diagonalization was capable of significantly outperforming both compressed sensing and super resolution because of its ability to sample off the grid. Even still, the technique was sensitive to a broad range of parameters, which were capable of making it divergent if chosen incorrectly. Given enough tuning and the appropriate signal form, however, filter diagonalization is the superior method for these types of signals.


%

\section{Acknowledgements}
We acknowledge the financial support of Defense Advanced Research Projects Agency grant N66001-10-1-4063 and the Defense Threat Reduction Agency under contract no. HDTRA1-10-1-0046. T.M. acknowledges support from the National Science Foundation (NSF) through the Graduate Research Fellowship Program (GRFP).  J.N.S. acknowledges support from the Department of Defense (DoD) through the National Defense Science {\&} Engineering Graduate Fellowship (NDSEG) Program. S.B. acknowledges support from the Department of Energy (DoE) through the Computational Sciences Graduate Fellowship (CSGF). A.A.G. thanks the Corning Foundation. We acknowledge Professor Vladimir Mandelshtam for useful conversations.

\end{document}